\documentclass[lettersize,journal]{IEEEtran}
\usepackage{amsmath,amsfonts}
\usepackage{algorithmic}
\usepackage{array}
\usepackage[caption=false,font=normalsize,labelfont=sf,textfont=sf]{subfig}
\usepackage{textcomp}
\usepackage{stfloats}
\usepackage{url}
\usepackage{verbatim}
\usepackage{graphicx}
\usepackage[T1]{fontenc}
\usepackage{balance}
\usepackage[backend=biber,style=ieee,sorting=none]{biblatex}
\usepackage[hidelinks,colorlinks=true,linkcolor=blue,citecolor=blue]{hyperref} 
\usepackage{multicol}
\usepackage{graphicx}
\usepackage{multirow} 
\usepackage{makecell} 
\usepackage{xcolor} 
\usepackage{soul} 
\definecolor{nice_blue}{RGB}{0, 102, 204}   
\definecolor{nice_red}{RGB}{204, 0, 0}      
\def\BibTeX{{\rm B\kern-.05em{\sc i\kern-.025em b}\kern-.08em
    T\kern-.1667em\lower.7ex\hbox{E}\kern-.125emX}}
\begin{document}
\title{Performance characterization of a new Structural and Thermal
Architecture for a future spaceborne Closed-Cycle Dilution Refrigerator}
\author{Valentin Sauvage$^a$,
	Anaïs Besnard$^a$,
	Clémence de Jabrun$^a$,
	Mehdi Bouzit$^a$,
	Bruno Maffei$^a$\\
	$^a$ IAS, Université Paris-Saclay, CNRS, Institut d’Astrophysique Spatiale, 91400, Orsay, France.
	\thanks{Contact: valentin.sauvage@universite-paris-saclay.fr}}
\maketitle
\begin{abstract}
	A Structural and Thermal Model (STM) has been developed to support the new spaceborne Closed-Cycle Dilution Refrigerator (CCDR), which aims to provide continuous cooling at 100~mK for long-duration astrophysical missions. The STM is based on a hexapod architecture that ensures both thermal decoupling and mechanical robustness during launch. In this paper, we present the characterization of its thermal and mechanical performances. A dedicated experimental setup was used to investigate the thermal behavior of the STM across a broad temperature range. The study reveals limitations of the collar design, with incomplete power interception from thermal boundary resistances and vibration test failure traced to defective strut gluing. These results guide the next STM iteration with optimized collar and strut assembly for reliable CCDR operation in space.
\end{abstract}
\section{Introduction}
\IEEEPARstart{E}{xperimental} cosmology from space demands ever-innovative and high-performance technologies to enable longer and more sensitive observations. Cooling systems often impose limitations on the mission lifetime of the spacecraft and on the thermal stability of sub-Kelvin detection chains. To address these challenges, a \textbf{C}losed-\textbf{C}ycle \textbf{D}ilution \textbf{R}efrigerator (CCDR) designed for space applications is under development~\cite{martin_closed_2010,volpe_developpement_2014,sauvage_development_2023}. The CCDR provides continuous cooling at ultra-low temperatures (100 mK) using helium isotopes $^3$He and $^4$He. Unlike the \textbf{O}pen-\textbf{C}ycle \textbf{D}ilution \textbf{R}efrigerator (OCDR) used in missions such as Planck-HFI~\cite{triqueneaux_design_2006}, the CCDR does not deplete helium resources, enabling long-duration missions. \\
The cooling power of the CCDR is generated by mixing $^3$He and $^4$He, which are then separated and re-circulated within the system. This design offers significant advantages over the OCDR, including higher cooling power (>~2~$\mu$W compared to <~0.2~$\mu$W), extended observation periods (>~3 years versus <~2.5 years), and equivalent temperature stability (20~nK·Hz$^{0.5}$). The CCDR is tailored for microgravity environment, making use of a porous sponge to separate liquid and gas phases in the "still", a circulator to pump the vapor phase, mainly composed of $^3$He, and a fountain pump to extract superfluid $^4$He, ensuring continuous cooling~\cite{martin_closed_2010,volpe_developpement_2014,sauvage_development_2023}. Laboratory testing has demonstrated the system feasibility, achieving \textbf{T}echnology \textbf{R}eadiness \textbf{L}evel (TRL) 4. Advancing to TRL 5 requires further validation through an \textbf{E}ngineering \textbf{M}odel (EM). \\
A hexapod-based Structural and Thermal Model (STM) has been developed to house the CCDR subsystems, to support the future focal plane at sub-K temperature, while interfacing with the 1.7~K stage.
It therefore needs to ensure thermal decoupling between thermal stages, and to withstand the intense vibrational stresses experienced during launch. This STM has been developed using the large heritage from past (Planck) and current missions (Athena). The thermal performance of the first version of the STM has been evaluated, showing promising results~\cite{sauvage_new_2022,sauvage_development_2023}, leading to further work reported in this paper.
\section{Status of the STM}
\label{sec:status_previous}
The purpose of the CCDR STM (Fig.~\ref{fig:overview_full_stm}) is to thermally decouple the coldest stage (cold plate) from the other components of the system. To achieve this, it is essential to estimate the amount of heat transferred through the struts, from the various sub-systems to the cold plate where the cooling power of the dilution is produced. The greater this thermal load is, the more limited the remaining cooling power for the future focal plane will be. For this reason, an experimental setup was developed to measure this heat transfer as a function of the temperature of the various sub-systems. This aspect is addressed in Section~\ref{sec:thermal_performances}. The mechanical aspects must also be evaluated. The STM must withstand the vibrations experienced during launch to ensure proper operation once in space. The complete setup used to assess the mechanical performance, along with the corresponding results, is presented in Section~\ref{sec:mechanical_performances}.
\subsection{STM description}
\begin{figure*}
	\includegraphics[width=\textwidth]{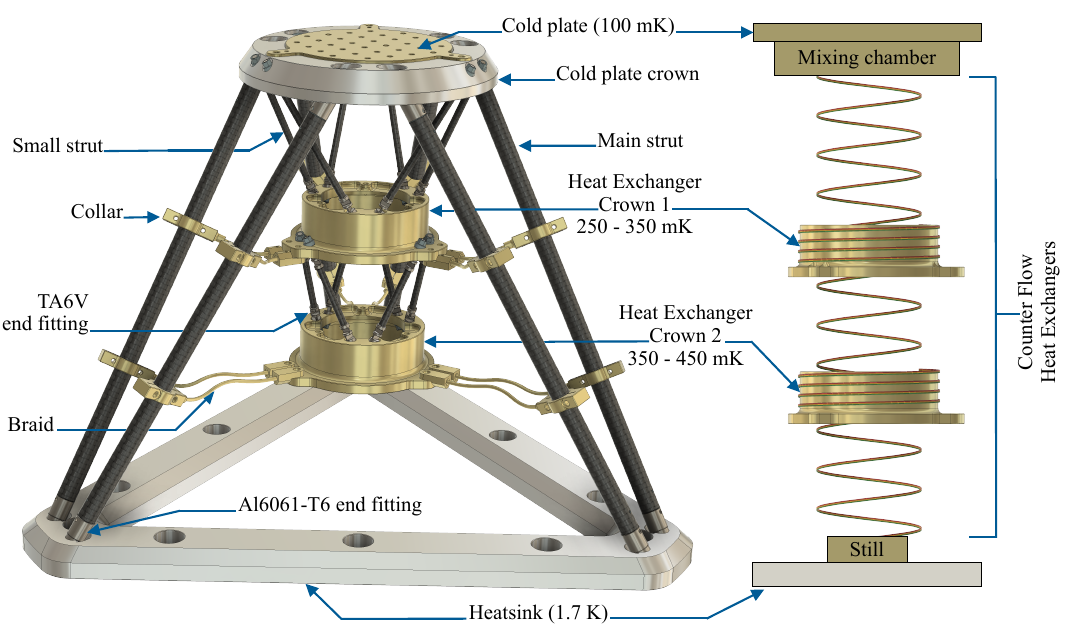}
	\caption{ CAD model of the CCDR STM. The cold plate is mounted within the cold plate crown. The heat sink stage and the cold plate crown are connected via the main struts. Two additional sets of smaller struts support the Heat Exchanger Crowns (HECs): HEC1, operating in the 250–350~mK range, and HEC2, in the 350–450~mK range. HEC1 and HEC2 are thermally linked to the main struts using collars and copper braids.}
	\label{fig:overview_full_stm}
\end{figure*}

The CCDR STM (Fig.~\ref{fig:overview_full_stm}) has been extensively described in previous work~\cite{sauvage_development_2023,sauvage_new_2022} and only a brief summary is provided here. The cold plate is made of gold-plated OFHC\footnote{\textbf{O}xygen-\textbf{F}ree \textbf{H}igh \textbf{C}onductivity copper, known for its high purity and excellent thermal conductivity} copper, and is positioned directly above the dilution refrigerator’s mixing chamber, as illustrated schematically on the right of Fig.~\ref{fig:overview_full_stm}. A future focal plane will be mounted on top of this cold plate. The cold plate is supported by a superconducting crown made of Al6061-T6, which is mechanically and thermally connected to the Al6061-T6 heat sink, serving as the interface with the rest of the cryogenic chain. These two components are connected via CFRP T700 struts, each equipped with dedicated Al6061-T6 end fittings. Each main strut includes two thermal intercept collars that reduce heat conduction from the heat sink to the cold plate. These collars are thermally linked to the two \textbf{H}eat \textbf{E}xchanger \textbf{C}rowns (HECs) using gold-plated copper braids. The HECs are cooled by the residual cooling power of the dilution refrigerator and are suspended in the center of the structure using small CFRP T700 struts with TA6V end fittings, as represented in Fig.~\ref{fig:overview_full_stm}.
\subsection{Previous steps}
A first version of the STM was thermally qualified. This initial configuration consisted of a reduced structure, comprising only the cold plate, the cold plate crown, the main struts, and the heat sink. Thermal characterization was carried out on this version. Throughout the paper, we often refer to the temperature pair 100~mK (cold plate) / 1.7~K (heat sink), which represents a typical baseline for numerous space missions. The full experimental setup and procedure are detailed in~\cite{sauvage_development_2023}.
The measured heat load on the cold plate was $\dot{Q} = 2.7~\mu$W, which is significantly lower than the predicted value of $\dot{Q} = 7.9~\mu$W. This discrepancy arises because the thermal model only accounts for bulk thermal conductivity. This model does not include the effects of multiple thermal interfaces, in particular those at the main strut end fittings (e.g., Kapitza resistance), the superconducting state of the cold plate crown, and the limited thermal coupling between the cold plate and its crown. The addition of the Heat Exchanger Crowns, coupled via thermal collars to intercept the heat conducted through the main struts, is expected to substantially reduce the value of $\dot{Q}$ from 2.7~$\mu$W (without HECs) to below 0.63~$\mu$W (with HECs). This model does not include the capillaries between the HECs and the cold plate, which would reduce the net available cooling power (more details on model: \cite{sauvage_development_2023}, p.226). The present study focuses solely on the thermo-mechanical performance of this STM.
\section{Characterization of the thermal performances}
\label{sec:thermal_performances}
For the thermal performance characterization, the STM was placed within a test cryostat (Bluefors LD400) to provide the necessary temperatures stages.
\subsection{The additional braids for specific tests}
\begin{figure}[h]
	\includegraphics[width=0.5\textwidth]{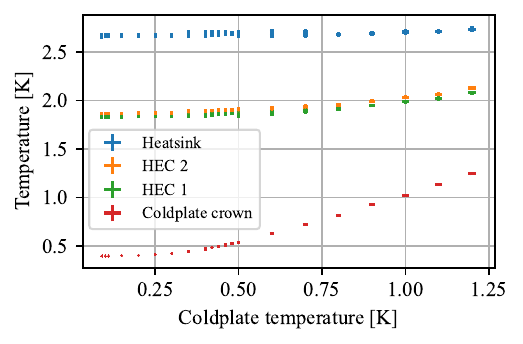}
	\caption{The minimum temperatures reached by the various stages of the CCDR are compared to the cold plate temperature. These differences are primarily attributed to the weak thermal coupling between the cold plate and the different components of the CCDR.}
	\label{fig:thermal_decoupling_without_braid}
\end{figure}

\begin{figure}[h]
	\includegraphics[width=0.5\textwidth]{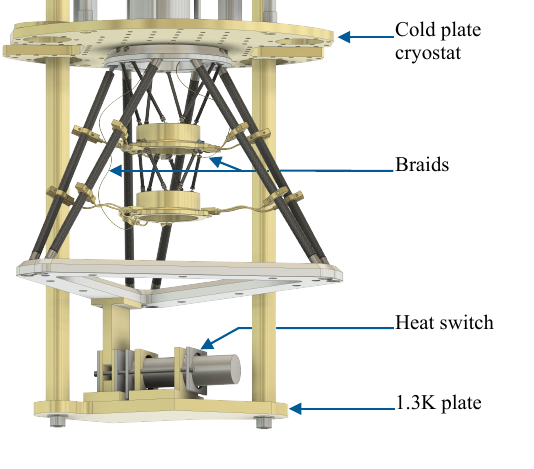}
	\caption{CAD view of the STM assembly, including the mechanical heat switch used to facilitate the cool down of the structure inside the DRACuLA cryostat at IAS (Bluefors model LD400). Two copper braids connect the Heat Exchanger Crowns to the STM cold plate, enabling improved thermal coupling and more accurate characterization of the structure at very low temperatures.}
	\label{fig:view_stm_inside_dracula}
\end{figure}

\begin{figure}[h]
	\includegraphics[width=0.5\textwidth]{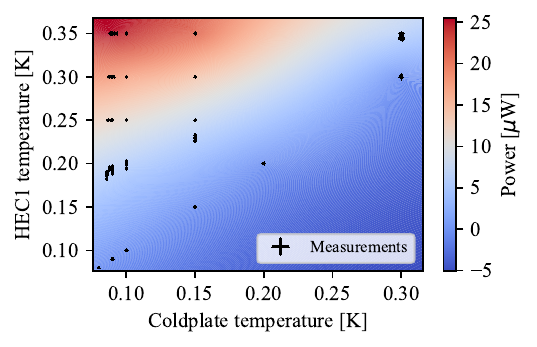}
	\caption{The thermal conduction of \textit{braid 1}, which connects Heat Exchanger Crown 1 to the CCDR cold plate and assists in the cooling of the HEC, has been characterized. Several measurement points were used to determine the thermal properties of the braid (black dots). A quadratic estimator was applied to explore different temperature combinations (shown in color).}
	\label{fig:braid1_thermal_conduction}
\end{figure}

Because one of the main functions of this structure is to ensure thermal decoupling between the different elements, passive cooling to very low temperatures (below 200~mK) cannot be achieved. As illustrated in Fig.~\ref{fig:thermal_decoupling_without_braid}, the minimum temperature passively reached by the various stages (Y-axis), while the STM cold plate is cooled to a given temperature (X-axis), remains insufficient for carrying out the thermal characterization of the structure. This limitation originates primarily from the weak thermal coupling between the cold plate (used to cool the STM) and the cold plate crown. Later, once the dilution unit is integrated into this structure, we expect to be able to extract $\approx~15~\mu W$ at 300~mK (HEC1) and $\approx~40~\mu W$ at 400~mK (HEC2). The braids are used to emulate this anticipated cooling power. This anticipated cooling power originates from the residual internal energy still carried by the circulating $^3$He--$^4$He mixture at the HEC stages, before it reaches the mixing chamber. Following the same enthalpy-balance approach used to derive the mixing-chamber cooling power ($\dot{Q}(T) \propto \dot{n}_4 T^2$, see~\cite{sauvage_development_2023}, eq. 5.17), the mixture retains a fraction of its cooling capacity at any intermediate temperature along the counter-flow heat exchanger, which can be intercepted at the HEC stages instead of being carried further to 100~mK. For the nominal design flow rates of the CCDR-EM ($\dot{n}_3 = 30~\mu$mol.s$^{-1}$, $\dot{n}_4 = 210~\mu$mol.s$^{-1}$), this yields cooling power values of the order of those quoted above at 300~mK and 400~mK. Additional details on the CFHX model and the flow rates can be found in~\cite{sauvage_development_2023}, chapters 5 and 10. A complete, dedicated derivation and validation of the CFHX sizing model, including its application to the HEC1/HEC2 cooling power estimates, will be presented in a forthcoming paper focused specifically on the CFHX design.
Additional details can be found in~\cite{sauvage_development_2023}.
Therefore, for these tests, additional copper braids (see Fig.~\ref{fig:view_stm_inside_dracula}) were used to thermally connect the cold plate to Heat Exchanger Crown 1 (braid~1) and Heat Exchanger Crown 2 (braid~2). A2-70 M2.5×12 screws were used for assembly, with a torque of 0.57~N·m, taking advantage of the presence of stainless-steel inserts in the threads of both the cold plate and the heat exchanger components. Temperature monitoring is performed using Lakeshore RX-102A on the heat exchanger side and Lakeshore RX-102B on the cold plate side.
Because the braids used to help the cool down conduct a large part of the heat, their effect must be measured for each temperature setting to be correctly included in the analysis. To do this, after the main experiment, the structure was partly dismantled inside the cryostat: the braids were disconnected from the heat exchangers (HECs) but left attached to the cold plate. The cold plate interface was left untouched, since disturbing it at low temperature could modify the thermal contact and introduce unquantifiable errors. The thermometers and heaters that were first installed on the HECs were then connected to the free ends of the braids. The same sequence of temperature steps as in the main experiment was repeated, and the injected power was recorded, giving the results in Fig.~\ref{fig:braid1_thermal_conduction}. A quadratic fit was then used to interpolate and extrapolate the data, so the power conduction could be estimated over a wider range of temperatures.
The measured thermal conduction of braid~1 ranges from $5.2~\mu$W to $25.2~\mu$W over the 83 measured points, spanning $T_\mathrm{HEC1}=182$--$350$~mK and $T_\mathrm{coldplate}=86$--$301$~mK, consistent with the expected conductance of a gold-plated copper braid of this cross-section and length at sub-Kelvin temperatures. This range already covers the full parameter space explored in the main characterization (Section~\ref{sec:thermal_performances}-D: coldplate 90--300~mK, HEC1 250--350~mK), so the quadratic estimator is used here mainly for interpolation rather than extrapolation. A two-dimensional quadratic fit $\dot{Q}(T_\mathrm{HEC1},T_\mathrm{coldplate})$ captures the data with $R^2=0.9999$ and a residual scatter of $0.6\%$ of the measured power (reduced $\chi^2\approx1.3$ against the $\approx1\%$ relative measurement uncertainty), supporting the use of the quadratic estimator for interpolation. The leading coefficient of the fit, $\approx2.2\times10^{-4}$~W.K$^{-2}$, is compatible with a simplified model $\dot{Q}\propto T_\mathrm{HEC1}^2-T_\mathrm{coldplate}^2$ ($R^2=0.995$), consistent with the residual electronic (Wiedemann-Franz-like) contribution to the thermal conductivity of copper in this temperature range. Braid~2, connecting HEC2 to the cold plate, was characterized following the same method and yields consistent results within its own temperature domain ($T_\mathrm{HEC2}=313$--$475$~mK), with a comparable fit quality and a leading coefficient of the same order of magnitude.
\subsection{Experimental setup inside the test cryostat}
During the characterization of the first version of this STM, a gas-gap heat switch was used to assist the cool down of the heat sink. However, this introduced additional power losses in the form of parasitic heat (few $\mu$W during operation, more details on \cite{sauvage_development_2023}, p.217), along with associated uncertainties. To eliminate this effect, a mechanical heat switch—provided by Entropy GmbH—was selected for the current setup, ensuring no parasitic heat during operation ($\dot{Q}$ = 0~$\mu$W). The heat switch was mounted on a support plate thermally anchored to the still plate of the cryostat via two copper rods. During the experiment, this plate was passively maintained at a temperature of 1.3~K. When activated, the heat switch establishes a thermal connection between the STM heat sink and the support plate. The radiative environment of the experiment was maintained at 3~K, provided by a radiation shield thermally anchored to the second stage of the cryostat’s pulse-tube cooler.
The STM (the test object) is directly and rigidly anchored, via its own cold plate, to the mixing chamber plate of the Bluefors LD400 cryostat. In practice, the STM cold plate temperature is the regulated quantity, stabilized by closed-loop heating between 90~mK and 300~mK over the course of the experiment (see Section~\ref{sec:thermal_performances}-D for the complete set of temperature steps); the cryostat's own mixing chamber plate is left uncontrolled and passively cooled to its base temperature, providing the ultimate heat sink for the STM cold plate. The STM itself is suspended in an inverted configuration below the still plate of the cryostat, with the heat sink supported by the mechanical heat switch bracket, as shown in Fig.~\ref{fig:view_stm_inside_dracula}.
\subsection{Instrumentation}
Multiple thermometers and heaters were used to accurately characterize the thermal behavior of the STM. TABLE~\ref{tab:instrumentation} summarizes the relevant instrumentation used as well as the equipment for their readout. The thermometers used in this study were calibrated against the primary ruthenium oxide thermometer installed on the cold plate of the Bluefors cryostat. A 4-wires measurement technique was employed to read the sensors, with excitation voltages selected according to Lakeshore’s recommendations to ensure accuracy. The heaters consist of thin metallic film resistors housed in custom-made, gold-plated copper enclosures. These are also connected using a 4-wires configuration to allow precise determination of the injected power.
\begin{table}[]
	\caption{\textcolor{nice_red}{Heaters} and \textcolor{nice_blue}{Thermometers} used as instrumentation.}
	\begin{tabular}{|p{2cm}|p{2.5cm}|p{3cm}|}
		\hline
		                           & Instruments                              & Read-out equipments        \\ \hline
		\multirow{2}{*}{Coldplate} & \textcolor{nice_blue}{Lakeshore RX-102B} & Lakeshore 372              \\ \cline{2-3}
		                           & \textcolor{nice_red}{10~k$\Omega$}       & \makecell{V: Fluke 8842a   \\I: Lakeshore 372}                                 \\ \hline
		Coldplate crown            & \textcolor{nice_blue}{Lakeshore RX-102B} & Lakeshore 370              \\ \hline
		\multirow{2}{*}{HEC1}      & \textcolor{nice_blue}{Lakeshore RX-102A} & Cryo-con Model 54          \\ \cline{2-3}
		                           & \textcolor{nice_red}{2~k$\Omega$}        & I + V: Keithley 2602A      \\ \hline
		\multirow{2}{*}{HEC2}      & \textcolor{nice_blue}{Lakeshore RX-102A} & Cryo-con Model 54          \\ \cline{2-3}
		                           & \textcolor{nice_red}{3~k$\Omega$}        & I + V: Keithley 2602A      \\ \hline
		\multirow{2}{*}{Heatsink}  & \textcolor{nice_blue}{Lakeshore CX-1010} & Lakeshore 336              \\ \cline{2-3}
		                           & \textcolor{nice_red}{10~k$\Omega$}       & \makecell{V: Yokogawa 7562 \\I: keithley 2000}                               \\ \hline
	\end{tabular}
	\label{tab:instrumentation}
\end{table}
\subsection{Procedure and analysis}
A broad temperature range was explored during the experiment (Coldplate: lowest, 90, 100, 150 and 300~mK; HEC1: lowest, 250, 300 and 350~mK; HEC2: lowest, 350, 400 and 450~mK; Heatsink: lowest, 1.6, 1.7 and 1.8~K), resulting in 320 distinct temperature combinations. At each point, the injected power was monitored across all four stages using a closed-loop feedback system to stabilize the set temperatures. Because the sub-systems of the STM do not naturally reach the cold plate temperature, this leads to the presence of a parasitic heat load, denoted as $\dot{Q}_0$, which constitutes the primary source of uncertainty in this study.
It is essential to investigate the behavior of the structure when its components are at their minimum achievable temperatures. For example, the heat sink passively reaches a minimum temperature of 1.28~K, indicating that a parasitic heat flow $\dot{Q}_0$ is conducted through the main structural struts. To maintain the heat sink at a higher temperature of 1.7~K, a controlled heating power $\dot{Q}_\textrm{heatsink}$ must be applied. Consequently, the total power required to stabilize the heat sink at 1.7~K is the sum $\dot{Q}_\textrm{heatsink} + \dot{Q}_0$.
The value of $\dot{Q}_0$ can be estimated from the measured thermal conduction through the braid, corresponding to the power required to sustain a temperature gradient between the cold plate and the minimum achievable temperature of Heat Exchanger Crown 2. This estimation assumes that the thermal intercept collars capture the entirety of the parasitic heat flow. In other words, that the temperature of the CFRP struts at the collars level is equal to that of the collars, which is also discussed later.
\subsection{Results of the thermal characterization}
\begin{figure}[h]
	\includegraphics[width=0.5\textwidth]{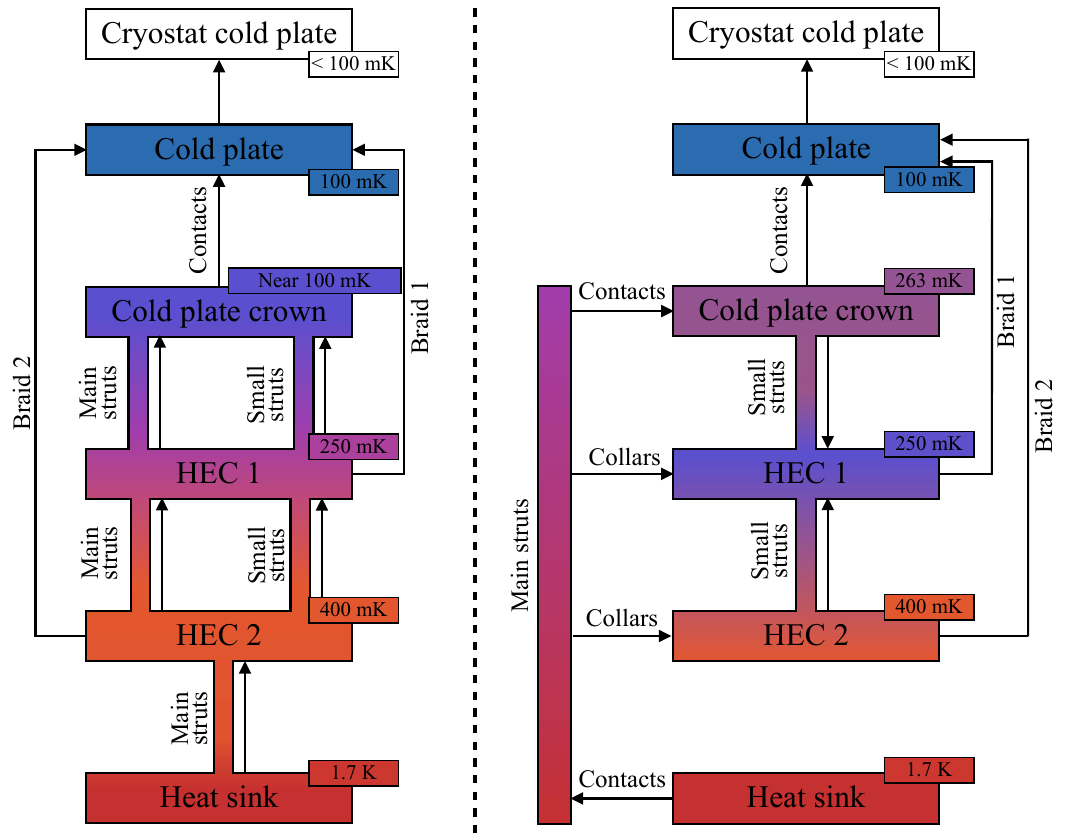}
    \caption{Schematical comparison of the heat circulation in the structure: expected case (left) and observed case (right). Both panels are illustrated for the same set point configuration, chosen within the explored temperatures range as a representative operating point at which the deviation between the ideal expectation (cold plate crown near 100~mK) and the actual observed behavior (cold plate crown at 263~mK) is clearly visible, illustrating the central finding of Section~\ref{sec:thermal_performances}-E.}
	\label{fig:schematic_thermal_path}
\end{figure}

In the ideal case, assuming the thermal intercept collars capture the entirety of the parasitic heat flow (i.e., perfect braids with zero temperature drop, and CFRP fully thermalized at the collar level), the power balance of Fig.~\ref{fig:schematic_thermal_path}-left can be decomposed node by node: $\dot{Q}_1 \approx 35.8~\mu$W from the heat sink to the HEC2 collars, $\dot{Q}_2 \approx 0.49~\mu$W from the HEC2 collars to the HEC1 collars, $\dot{Q}_3 \approx 0.50~\mu$W from the HEC1 collars to the cold plate crown, $\dot{Q}_4 \approx 0.27~\mu$W from HEC2 to HEC1 via the small struts, and $\dot{Q}_5 \approx 0.13~\mu$W from HEC1 to the cold plate crown via the small struts, giving a total of $\dot{Q}_3+\dot{Q}_5 \approx 0.63~\mu$W reaching the cold plate crown, consistent with the value quoted in Section~\ref{sec:status_previous}-B. Because this residual power is small compared to $\dot{Q}_1$, the cold plate crown remains close to the cold plate temperature in the ideal case, labeled as ``near 100~mK'' on Fig.~\ref{fig:schematic_thermal_path}-left.
The method described above, especially the role of the braids, shows the impact of thermal boundary resistance at the collar. Although perfect heat interception by the collars was expected (see Fig.\ref{fig:schematic_thermal_path}, left), adding the braids forces heat to pass through the CFRP/collar interface. This interface introduces a thermal boundary resistance with a temperature dependency. At equilibrium, the temperature difference $\Delta T$ across the interface is proportional to both the interfacial resistance $R(T)$ and the heat flow $\dot{Q}$ through the contact area $A$. Since $R(T)$ also depends on the contact quality and increases as temperature decreases ($R(T)\propto T^{-n}$ with $n>1$)~\cite{pobell_matter_2007}, the effect only appears once the braids are connected. \\
The consequence of this is that, since the collars design is not efficient to intercept the heat in this temperature range (meaning that the CFRP at the collar level is not well thermalized), the power balance becomes more difficult to establish. A part of the heat is not intercepted by the HEC2 collars, goes to the HEC1 collars, is again not intercepted, and finally flows directly into the cold plate crown. This is our interpretation of the fact that, at the passive (unheated) HEC1/HEC2 operating point, HEC1 is measured to be colder than the cold plate crown, leading to the diagram in Fig.~\ref{fig:schematic_thermal_path}-right.
At the baseline point (coldplate 100~mK, heat sink 1.7~K, HEC1 and HEC2 unheated), the measured temperatures are $T_\mathrm{coldplate\,crown}=261.8$~mK, $T_\mathrm{HEC1}=199.0$~mK and $T_\mathrm{HEC2}=380.1$~mK, i.e., HEC1 is $63$~mK colder than the cold plate crown. Repeating this measurement at fixed heat sink (1.7~K) while sweeping the cold plate set point from 85 to 300~mK shows that the gap $\Delta T=T_\mathrm{coldplate\,crown}-T_\mathrm{HEC1}$ decreases monotonically, from $\approx72$~mK at 85~mK to $\approx39$~mK at 150~mK, down to $\approx-8$~mK (i.e., HEC1 and the crown become equal within measurement uncertainty) at 300~mK. This trend is a direct experimental signature of the temperature-dependent interfacial resistance $R(T)\propto T^{-n}$ invoked above: at lower absolute temperature, $R(T)$ is larger, so the HEC1 collar intercepts proportionally less of the heat flowing down the main strut, leaving a larger fraction to continue into the cold plate crown; as the temperature increases, $R(T)$ decreases, interception becomes more effective, and the gap closes accordingly. By adding the parasitic heat, which also increases the complexity of the power diagram, the structure becomes too complex without additional characterization on the copper/CFRP interface.
\subsection{Conclusion on thermal performances}
The thermal characterization method developed here highlights that the power interception on the main struts is not sufficient enough for the objective of this structure. We propose splitting the main struts at the collar level: both tubes would be bonded to an intermediate fitting (similar to those used at both ends), and the braids would be terminated and connected at this interface. This configuration should provide complete heat interception and introduce additional thermal boundary resistance, enhancing thermal decoupling, but it may reduce STM stiffness. Once this redesign is completed, the same method will be applied to assess the thermal performance of the new STM.
The choice of the main strut material was the subject of a dedicated trade-off study prior to this work~\cite{sauvage_development_2023}, comparing CFRP with niobium-titanium (NbTi), Al6061-T6 and stainless steel (SS316) over the full 100~mK--1.7~K span. While Al6061-T6 shows a favorable thermal conductivity near 100~mK, it becomes significantly more conductive near 1.7~K, making it unsuitable across the full temperature range. NbTi and SS316 are both more thermally conductive than CFRP over this range. CFRP was retained because it offers the lowest thermal conductivity of the compared materials while allowing the mechanical stiffness to be tuned, to a limited extent, through the fiber orientation and layup, via the same trade-off between the $S/L$ ratio and the second moment of area used to size the current struts. The present results indicate that the limiting factor is not the bulk conductivity of the CFRP strut itself, but the thermal boundary resistance at the CFRP/collar interface. This is why the proposed redesign targets the collar/strut interface rather than a change of the bulk strut material.
\section{Mechanical performances}
\label{sec:mechanical_performances}
\subsection{Protocol}
\label{subsec:protocol_mechanic}
To evaluate the mechanical performance of the structure, a standard vibration test protocol was conducted using the vibration test facility at IAS is listed below.
\begin{itemize}
\item Low-level sine: Initial sweep to identify resonances.
\item High-level sine: Stronger sweep to test structure.
\item Low-level sine: Control sweep.
\item Random -12~dB: Reduced random vibration.
\item Low-level sine: Control sweep.
\item Random -6~dB: Intermediate random vibration.
\item Low-level sine: Control sweep.
\item Random 0~dB: Full-level qualification random vibration.
\item Low-level sine: Final control sweep.
\end{itemize}
The objective was to ensure that the first structural mode exceeds 140~Hz (first amplification frequency), with no visible damage and no change in tightening torques after the test. The mechanical load was gradually increased, and low-level sine sweeps were performed between excitation steps to detect any changes within the structure. A change of no more than 5\% in frequency and 20\% in amplitude before and after vibration tests was used as the acceptance criterion for structural integrity.
\subsection{Results}
\begin{figure}[h]
	\includegraphics[width=0.5\textwidth]{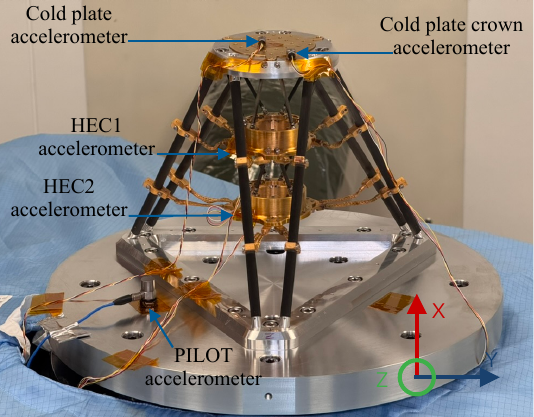}
	\caption{Photograph of the CCDR STM mounted on the IAS vibration facility, equipped with three-axis accelerometers on each stage (except the heat sink, which is mechanically coupled to the shaker bed): cold plate, cold plate crown, HEC1 and HEC2 accelerometers, with their locations indicated by arrows. A PILOT accelerometer mounted on the shaker bed, used as the control reference for the vibration test, is also indicated.}
	\label{fig:ccdr_on_shaking_pot}
\end{figure}

\begin{figure}[h]
	\includegraphics[width=0.5\textwidth]{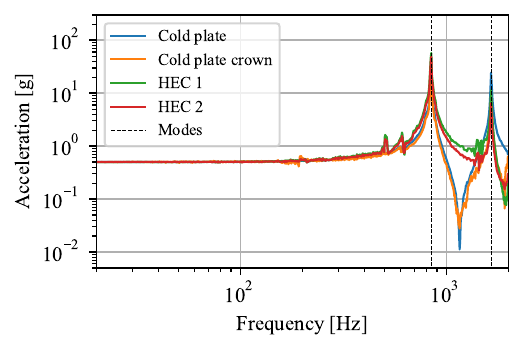}
	\caption{Acceleration spectra of the various structural elements as a function of frequency during the first low-level sine excitation along the X-axis. The first peak corresponds to the first mode of the structure.}
	\label{fig:shaking_spectra_example}
\end{figure}

\begin{figure}[h]
	\includegraphics[width=0.5\textwidth]{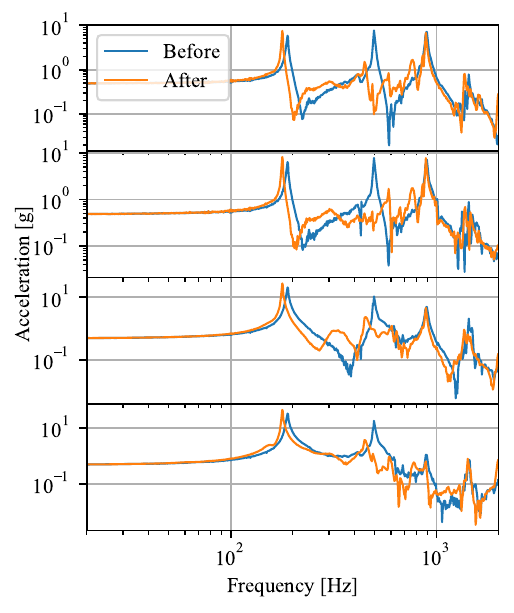}
	\caption{From top to bottom: cold plate, cold plate crown, HEC1 and HEC2. Comparison between the first and last low-level sine excitations on the Heat Exchanger Crown 1 (Y-axis), highlighting changes in the structural response in terms of both frequency and amplitude.}
	\label{fig:comparison_y_axis_spectra}
\end{figure}

The structure was vibrated along three axes. The X-axis corresponds to the vertical (top–bottom) direction, while the Y- and Z-axes are mutually perpendicular and both orthogonal to the X-axis, as illustrated in Fig.~\ref{fig:ccdr_on_shaking_pot}. Using accelerometers, we evaluate the displacement of the cold plate, the cold plate crown and the two Heat Exchanger Crowns.
Tri-axial 356A09 accelerometers from PCB Piezotronics were used on each stage to obtain the acceleration spectra. Fig.~\ref{fig:shaking_spectra_example} presents a typical acceleration spectrum obtained during the first low-level sine excitation along the X-axis. Distinct acceleration peaks can be observed, corresponding to the structural resonance modes, with the first peak representing the fundamental (first) mode of the structure.
By following the vibration test protocol previously defined, many spectra were obtained, which can then be compared one to the other. According to the criteria described in subsection~\ref{subsec:protocol_mechanic}, these results led to the pass/failed validation for the considered axis.. During the vibration tests of this structure in early May 2025, the X-axis was successfully validated (less than 0.2\% frequency shift and a maximum of 8.5\% in amplitude variation). However, the Y-axis did not pass the validation process. As shown in Fig.~\ref{fig:comparison_y_axis_spectra}, a significant difference between the first and the last low-level sine indicates a modification in the structure, leading to an investigation.
\subsection{Investigation}
\begin{figure}[h]
	\includegraphics[width=0.5\textwidth]{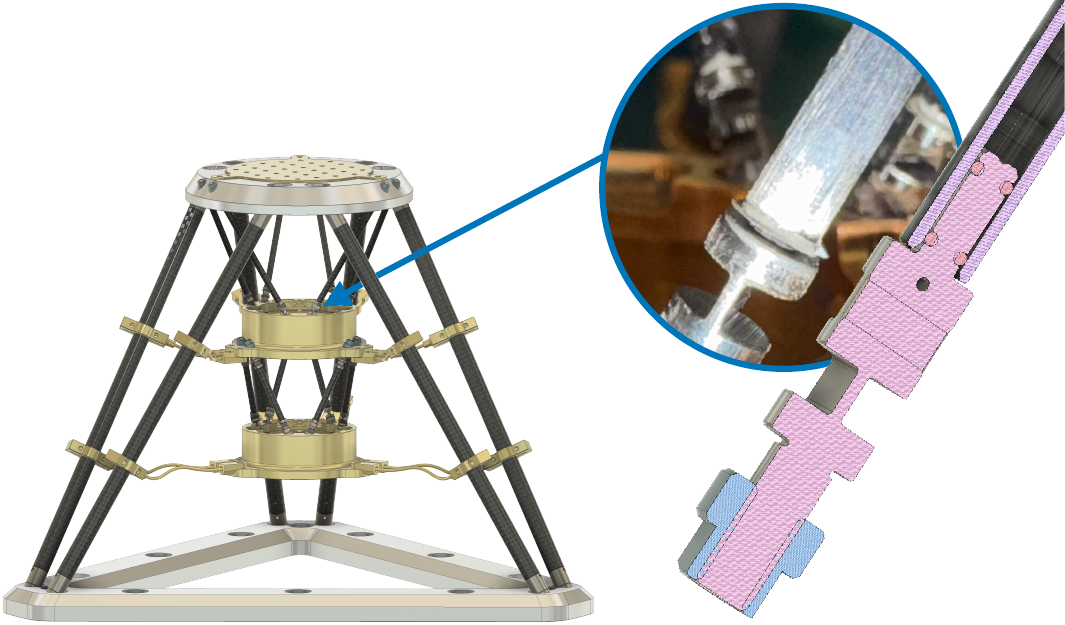}
	\caption{Picture of the broken element, showing its location in the structure, together with the CAD model of the small strut, focus on the end fitting area.}
	\label{fig:broken_part}
\end{figure}

A visual inspection quickly highlight a broken small strut, connecting the HEC1 to the cold plate crown. More precisely, the rupture was located at the connection between the TA6V end fitting and the small CFRP tube (visible in Fig.~\ref{fig:broken_part}), and was later attributed to a failure in the gluing process with Hysol9394. A 10~$\mu$m resolution X-ray tomography shown in Fig.~\ref{fig:tomography}, performed by RX SOLUTIONS in France, revealed that the gap between the end fitting and the CFRP tube was not completely filled with glue, leaving some zones without adhesive. This highlights the necessity to adapt our gluing process.\footnote{The process used for the main struts cannot be applied here, since the glue viscosity does not allow it, considering the very small dimensions involved.} The measurement of the Z-axis was not performed, since it was not relevant anymore.
\definecolor{cfrp_color}{RGB}{243,178,62}
\definecolor{fitting_color}{RGB}{130,215,83}
\definecolor{oring}{RGB}{72,160,249}
\definecolor{glue}{RGB}{219,61,38}
\begin{figure}[h]
	\includegraphics[width=0.5\textwidth]{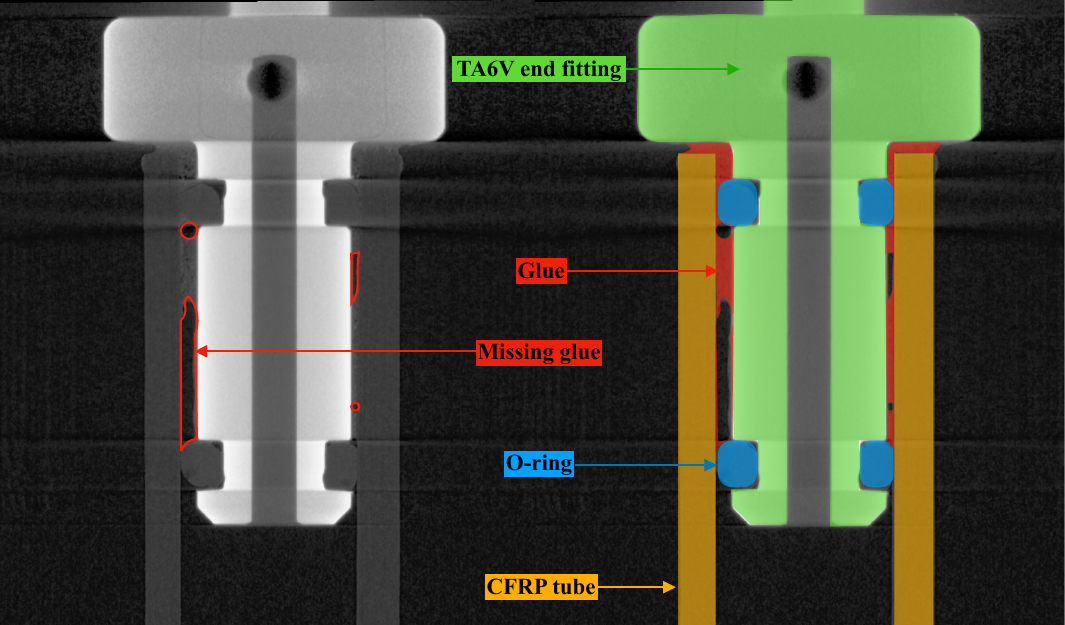}
	\caption{X-ray tomography of the TA6V strut at the end fitting level. \colorbox{fitting_color}{\textcolor{fitting_color}{test}} corresponds to the TA6V end fitting, \colorbox{cfrp_color}{\textcolor{cfrp_color}{test}} to the CFRP tube, \colorbox{oring}{\textcolor{oring}{test}} to the O-ring (ensuring glue containment), and \colorbox{glue}{\textcolor{glue}{test}} the glue . On the left is highlighted on red the volumes with missing glue.}
	\label{fig:tomography}
\end{figure}
We are actually investigating new techniques for this specific gluing, with the process to be validated by tomography. Afterwards, the new small struts will be integrated into the structure for further qualification tests.
\section*{Conclusion}
We reported on the performance characterization of the Structural and Thermal Model developed for the Closed-Cycle Dilution Refrigerator. The thermal study underlined that the present collars are not sufficient to fully intercept the parasitic heat flowing through the main struts, especially due to boundary resistances at the CFRP/collar interfaces. This result gives a clear guideline for the next design evolution, especially by increasing the interface surface and improving the thermal coupling. On the mechanical side, vibration tests confirmed that the structure is robust along the X-axis, but a small strut broke, so we could not validate the Y-axis and we could not repeat the thermal test to check any change in thermal performance. The failure was attributed to an incomplete gluing process, as confirmed by tomography. New gluing techniques are under development and will be validated before integration of the next generation struts. These improvements will be implemented in the future STM, which will be re-characterized both thermally and mechanically, paving the way toward a reliable Engineering Model of the CCDR.
\section*{Acknowledgments}
This work was supported by CNES, CNRS, the DIM ACAV programme from Île-de-France, Université Paris-Saclay, and OSUPS. The authors thank Bruno Crane for his assistance and RX Solutions for the tomography.
\printbibliography
\end{document}